\documentclass[prb,twocolumn,showpacs,superscriptaddress]{revtex4}%
\usepackage{amsfonts}
\usepackage{amsmath}
\usepackage{amssymb}
\usepackage{graphicx}%
\providecommand{\U}[1]{\protect\rule{.1in}{.1in}}

\begin{document}
\title{Thermodynamic phase transition of UH$_{3}$: Role of electronic strong correlation}
\author{Yu-Juan Zhang}
\affiliation{LCP, Institute of Applied Physics and Computational Mathematics, Beijing
100088, People's Republic of China}
\author{Bao-Tian Wang}
\affiliation{LCP, Institute of Applied Physics and Computational Mathematics, Beijing
100088, People's Republic of China}
\affiliation{Institute of Theoretical Physics and Department of Physics, Shanxi University,
Taiyuan 030006, People's Republic of China}
\author{Yong Lu}
\affiliation{LCP, Institute of Applied Physics and Computational Mathematics, Beijing
100088, People's Republic of China}
\author{Ping Zhang}
\thanks{Author to whom correspondence should be addressed. E-mail: zhang\_ping@iapcm.ac.cn}
\affiliation{LCP, Institute of Applied Physics and Computational Mathematics, Beijing
100088, People's Republic of China}
\affiliation{Center for Applied Physics and Technology, Peking University, Beijing 100871,
People's Republic of China}


\begin{abstract}
The electronic structure and thermodynamical properties of uranium trihydrides
($\alpha$-UH$_{3}$ and $\beta$-UH$_{3}$) have been studied using
first-principles density functional theory. We find that inclusion of strong
electronic correlation is crucial in successfully depicting the electronic
structure and thermodynamic phase stability of uranium hydrides. After turning
on the Hubbard parameter, the uranium 5\emph{f} states are divided into
well-resolved multiplets and their metallicity is weakened by downward shift
in energy, which prominently changes the hydrogen bond and its vibration
frequencies in the system. Without Coulomb repulsion, the experimentally
observed $\alpha\mathtt{\rightarrow}\beta$ phase transition cannot be
reproduced, whereas, by inclusion of the on-site correlation, we successfully
predict a transition temperature value of about 332 K, which is close to the
experimental result.

\end{abstract}
\maketitle

Since the report of UH$_{3}$ as one ferromagnetic (FM) uranium compounds
\cite{Trze1,Trze2,Troc}, it has attracted lots of attentions in nuclear fuel
field in the past decades. In solid state, uranium hydride exists mainly in
the form of cubic trihydride (Pm3n; No. 223) \cite{Bala} with two allotropes,
i.e., $\alpha$-UH$_{3}$ and $\beta$-UH$_{3}$. $\alpha$-UH$_{3}$ [Fig. 1(a)] is
an intermediate state from orthorhombic $\alpha$-U saturated with H to $\beta
$-UH$_{3}$ \cite{Tayl,Mulf,Geno}. It is the metastable low temperature phase
with two UH$_{3}$ formula units in one unit cell. The uranium and hydrogen
atoms in this structure occupy the 2(a) (0, 0, 0) and 6(c) (1/4, 0, 1/2)
sites, respectively. $\beta$-UH$_{3}$ [Fig. 1(b)] is the stable high
temperature phase with eight UH$_{3}$ formula units in one unit cell. The
uranium atoms in this structure occupy the 2(a) (0, 0, 0) and the 6(c) (1/4,
0, 1/2) sites, and the hydrogen atoms occupy the 24(k) (0, 0.156, 0.313)
sites. Mulford $et$ $al.$ \cite{Mulf} found through X-ray diffraction
measurement that the transition temperature of $\alpha$-UH$_{3}$ to $\beta
$-UH$_{3}$ is between 373 K and 523 K, while Genossar $et$ $al.$ \cite{Geno}
reported that $\alpha$-UH$_{3}$ transits irreversibly into $\beta$-UH$_{3}$
above room temperature. At even higher temperatures (above 673 K), uranium
hydride reversibly remove the hydrogen. This property makes uranium hydrides
convenient starting materials to create chemically reactive uranium powder
along with various uranium carbide, nitride, and halide compounds. In contrast
to the above mentioned experimental measurements on temperature-induced
$\alpha\mathtt{\rightarrow}\beta$ phase transition of uranium trihydride, to
date the theoretical exploration of this prominent phase transition is still
totally lacking in the literature. Considering the combined fact that (i)
UH$_{3}$ denotes a typical prototype among various kinds of actinide hydrides
and (ii) the actinide hydrides play extremely important role in nuclear fuel
design (as well as stockpile) in which the thermodynamic stability is a
substantial factor, therefore, a revealing theoretical study on the
ground-state properties and $\alpha\mathtt{\rightarrow}\beta$ phase transition
of UH$_{3}$ from first-principles quantum mechanics is highly needed for the
relevant important industrial applications.

\begin{figure}[ptb]
\begin{center}
\includegraphics[width=0.9\linewidth]{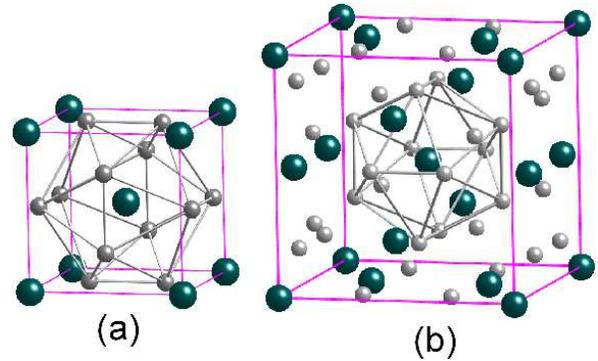}
\end{center}
\caption{(Color online) Crystal structures of (a) $\alpha$-UH$_{3}$ and (b)
$\beta$-UH$_{3}$, where the large and small balls denote the uranium and
hydrogen atoms, respectively.}%
\label{site}%
\end{figure}

From basic point of view, it can be visualized that many physical and chemical
properties of UH$_{3}$ are closely related to the quantum process of
localization and delocalization for partially filled uranium 5$f$ electrons.
Modeling of the electron localization/delocalization, and thus any
hydrogenation process involving uranium, is a complex task. Conventional
density functional theory (DFT) schemes that apply the local density
approximation (LDA) or the generalized gradient approximation (GGA)
underestimate the strong on-site Coulomb repulsion of the uranium 5$f$
electrons and consequently fail to capture the correlation-driven
localization. Therefore, the uranium 5$f$ electrons require special attention
in trying to gain any microscopic insight into the thermodynamical stability
of UH$_{3}$. Up to now several approaches, the LDA/GGA+\emph{U}, the hybrid
density functional of (Heyd, Scuseria, and Enzerhof) HSE, the self-interaction
corrected local spin-density (SIC-LSD), and the Dynamical Mean-Field Theory
(DMFT), have been developed to correct the pure LDA/GGA failures in
calculations of actinide compounds. Among these techniques, the effective
modification of pure DFT by LDA/GGA+\emph{U} formalisms has been confirmed
widely in study of uranium (as well as its neighbor, plutonium) oxides
\cite{Dudarev1,SunJCP,Zhan}. By tuning the effective Hubbard parameter in a
reasonable range, the antiferromagnetic (AFM) Mott insulator features of these
oxides were correctly calculated and the atomic structural parameters as well
as the electronic properties are well in accord with experiments. Inspired by
this series of successful calculations on AFM uranium oxide, as well as
motivated by the fact that uranium hydride is also a magnetic ordered (FM
instead of AFM) compound and thus a static mean-field treatment like
LDA/GGA+\emph{U} is expected to reliably catch its some key features in both
$\alpha$ and $\beta$ phases, in the present work, we investigate the
ground-state electronic properties and lattice dynamics of the two allotropes
of UH$_{3}$ using the LDA+\emph{U} formalism. For comparison the pure LDA
calculation is also performed. The most insightful result of our investigation
is that it is essential to take the 5$f$ on-site electronic correlation into
account for theoretically reproducing the experimentally observed
$\alpha\mathtt{\rightarrow}\beta$ phase transition of UH$_{3}$ at finite
temperature. Also, the prominent changes in the hydrogen bond and the
consequent hydrogen vibration (reflected by the optical branches in phonon
spectrum) in UH$_{3}$ by the inclusion of on-site Coulomb repulsion of the
uranium 5$f$ electrons is for the first time highlighted in this paper.

The calculations are performed using the projector-augmented wave (PAW) method
of Bl\"{o}chl \cite{paw}, as implemented in the Vienna \textit{ab initio}
simulation program (VASP) \cite{Kres}. For the plane-wave set, a cut-off
energy of 500 eV is used. The hydrogen 1\emph{s} and uranium 6\emph{s}$^{2}%
$6\emph{p}$^{6}$6\emph{d}$^{2}$5\emph{f}$^{2}$7\emph{s}$^{2}$ are treated as
valence electrons. The strong on-site Coulomb repulsion amongst the localized
uranium $5f$ electrons are accounted for by using the formalism formulated by
Dudarev \textit{et al}. \cite{Duda1}. In this scheme the total energy
functional is of the form%
\begin{equation}
E_{\text{LDA+}U}=E_{\text{LDA}}+\frac{U-J}{2}\sum_{\sigma}\left[
\text{Tr}\rho^{\sigma}-\text{Tr}\left(  \rho^{\sigma}\rho^{\sigma}\right)
\right]  , \label{e1}%
\end{equation}
where $\rho^{\sigma}$ is the density matrix of $f$ states, and $U$ and $J$ are
the spherically averaged screened Coulomb energy and the exchange energy,
respectively. Here the Coulomb $U$ is treated as a variable, while the
exchange energy is set to be a constant $J$=0.51 eV. This value of $J$ is in
the ball park of the commonly accepted one for uranium. Since only the
difference between $U$ and $J$ is significant \cite{Duda1}, thus we will
henceforth label them as one single parameter, for simplicity labeled as
\emph{U}$_{\mathrm{{eff}}}$, while keeping in mind that the non-zero $J$ has
been used during calculations. The Brillouin-zone (BZ) integrations are
performed using 9$\times$9$\times$9 and 7$\times$7$\times$7 Monkhorst-Pack
\cite{Monk} special \emph{k}-points for $\alpha$-UH$_{3}$ and $\beta$-UH$_{3}%
$, respectively. The geometries are optimized until the forces are less than
0.02 eV/\AA , and the total energy is relaxed until the difference value is
smaller than 10$^{-5}$ eV.

Through calculating the total energy dependences on \emph{U}$_{\mathrm{{eff}}%
}$ of $\alpha$-UH$_{3}$ and $\beta$-UH$_{3}$ in nonmagnetic, FM, and AFM
phases, we find that the FM phase is the most favorable state with
\emph{U}$_{\mathrm{{eff}}}$ from 0 to 7 eV, which is well consistent with the
experiment. The dependence of the lattice parameters on \emph{U}%
$_{\mathrm{{eff}}}$ for FM phase of $\alpha$-UH$_{3}$ and $\beta$-UH$_{3}$
demonstrates that in $\alpha$ phase, the lattice parameter is close to the
experimental value when \emph{U}$_{\mathrm{{eff}}}$ is in range of 4--6 eV,
while in $\beta$ phase, the range is in 4--5 eV. Considering the two
allotropes, therefore, we choose the value of \emph{U}$_{\mathrm{{eff}}}$ to
be 4 eV in our following study for the two phases. Our calculated lattice
constants for $\alpha$ ($\beta$) phase of UH$_{3}$ at \emph{U}$_{\mathrm{{eff}%
}}$=0 and 4 eV are 4.001 (6.382) and 4.128 (6.622) \r{A}, respectively.
Compared with pure LDA, our LDA+\emph{U} gives closer values with respect to
the experimental values of 4.160 \cite{Mulf} and 6.639 \cite{Rund1} \r{A}\ for
$\alpha$ and $\beta$ phase, respectively.

\begin{figure}[ptb]
\begin{center}
\includegraphics[width=1.0\linewidth]{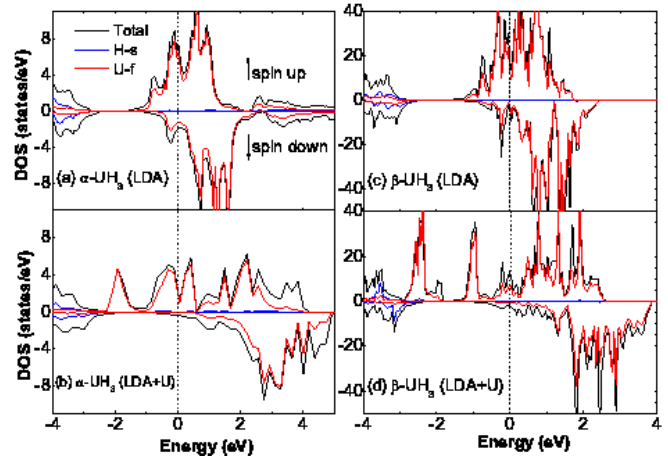}
\end{center}
\caption{(Color online) Spin-resolved total DOS for the two allotropes of
UH$_{3}$ calculated within LDA and LDA+\emph{U} (\emph{U}$_{\mathrm{{eff}}}$=4
eV) formalisms. Partial DOS of U 5\emph{f} and H 1\emph{s} orbitals are also
shown. The fermi level is set to be zero.}%
\label{dos}%
\end{figure}

Our calculated spin-resolved total density of states (DOS) and the partial DOS
for uranium 5$f$ and hydrogen 1$s$ states of the two allotropes are presented
in Fig. \ref{dos}. The upper panels show the pure LDA results, while the lower
ones give the LDA+\emph{U }results. From Fig. \ref{dos} we can observe evident
effects of strong correlation on the \emph{f} states. For both two phases of
UH$_{3}$, after turning on the Hubbard \emph{U} parameter, clearly, the
uranium 5\emph{f} states are divided into more well-resolved peaks compared
with pure LDA, and the energy distributions of these multiple 5\emph{f} peaks
are much localized and separated. As a result of this multiple peak splitting,
for $\alpha$ phase, the uranium 5\emph{f} states in the range of $-$1.0 to 2.0
eV within LDA are extended to lie in the range of $-$2.5 to 5.0 eV in the
LDA+\emph{U} formalism. For $\beta$ phase, similar effects of strong
correlation can be found. Moreover, the occupation of both spin-up and
spin-down electrons at the Fermi level are lowered after taking into account
the on-site Coulomb repulsion. However, no insulating band gap are opened by
the inclusion of Hubbard parameter and UH$_{3}$ still remains its metallic
nature as observed in the experiment \cite{Goud}.

Due to the decrease in metallicity of uranium 5$f$ states and their energy
downward shift towards the hydrogen 1$s$ orbital by taking account of the
electronic strong on-site correlation, the ionicity of the hydrogen bond is
prominently enhanced. To analyze the ionicity of the two allotropes of
UH$_{3}$, results from the Bader analysis \cite{Bader} are shown in Table
\ref{bader}. The charge (\emph{Q}) enclosed within the Bader volume (\emph{V})
is a good approximation to the total electronic charge of an atom. Note that
although we have included the core charge in charge density calculations,
since we do not expect variations as far as the trends are concerned, only the
valence charge is listed. From the results we calculated, it is seen that the
ionicity of the two allotropes within LDA+\emph{U} are more evident than that
within pure LDA. For $\alpha$ phase, the electrons transferred from each U
atom to H atoms are 1.473 and 1.575 within LDA and LDA+\emph{U} formalisms,
respectively. For $\beta$ phase, electrons transferred from two
inequipotential U atoms (U$_{\mathrm{{I}}}$ and U$_{\mathrm{{II}}}$) to H
atoms are 1.548 and 1.630, respectively, in LDA+\emph{U} formalism. Whereas,
in LDA formalism the values are 1.497 and 1.566 respectively.

\begin{table}[ptb]
\caption{Bader effective atomic charges and volumes of $\alpha$-UH$_{3}$ and
$\beta$-UH$_{3}$ in the LDA and LDA+\emph{U} (\emph{U}$_{\mathrm{{eff}}}$= 4
eV) formalisms.}%
\begin{ruledtabular}
\renewcommand{\tabcolsep}{0.01pc}
\begin{tabular}{cccccccccccccccc}
Allotrope&Methods &Q(U$_{\rm{I}}$)& Q(U$_{\rm{II}}$)& Q(H)& V(U$_{\rm{I}}$)&V(U$_{\rm{II}}$)&V(H)\\
\hline
$\alpha$-UH$_3$&LDA&12.527&&1.490&17.913&&4.703\\
&LDA+\emph{U}&12.425&&1.525&18.847&&5.415\\
$\beta$-UH$_3$&LDA&12.503&12.434&1.503&18.290&17.465&4.904\\
&LDA+\emph{U}&12.452&12.370&1.535&18.774&19.283&5.706\\
\end{tabular}
\label{bader}
\end{ruledtabular}
\end{table}

To illustrate the main point of this paper, i.e, the effect of electronic
strong on-site correlation on the thermodynamic properties of UH$_{3}$, we
have calculated the phonon dispersion and the Helmholtz free energy \emph{F}
of two allotropes with and without including Coulomb repulsion of 5\emph{f}
electrons. Phonon frequency calculations were carried out using the
Hellmann-Feynman theorem and the direct method \cite{Parl}. For $\alpha$ and
$\beta$ phases, we adopted 2$\times$2$\times$2 supercells containing 64 and
256 atoms with 3$\times$3$\times$3 and 1$\times$1$\times$1 Monkhorst-Pack
\emph{k}-point meshes in the BZ integration, respectively. The forces induced
by small displacements were calculated within VASP. The amplitude of all the
displacements is 0.03 \AA .

\begin{figure}[ptb]
\begin{center}
\includegraphics[width=8cm]{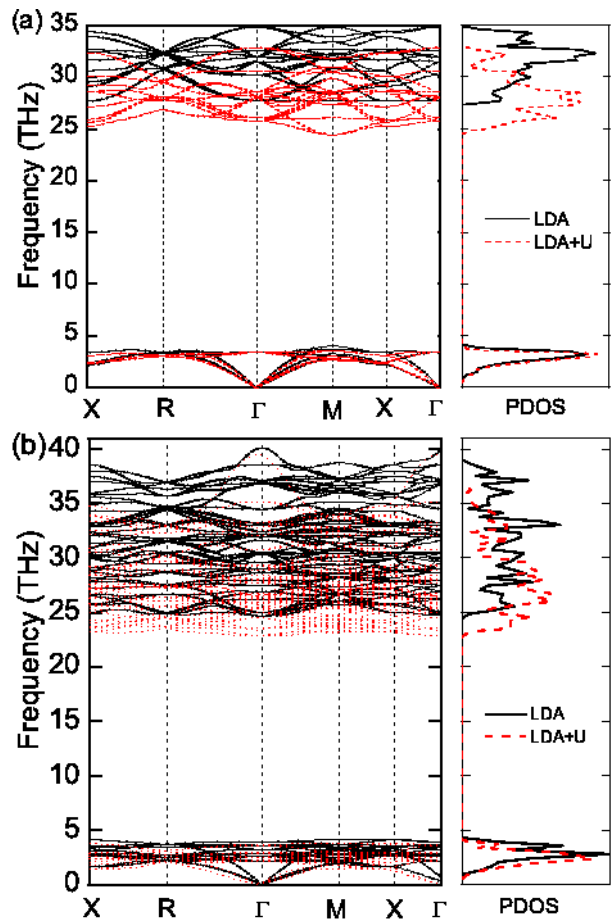}
\end{center}
\caption{(Color online) Phonon dispersion curves (left panel) and
corresponding phonon DOS (right panel) of (a) $\alpha$-UH$_{3}$ and (b)
$\beta$-UH$_{3}$. The black curves are computed within the LDA while the red
(grey) curves the LDA+\emph{U} formalism.}%
\label{phonon}%
\end{figure}

The calculated phonon curves along some high-symmetry directions in
the BZ together with the phonon DOS are plotted in Fig. 3. Clearly,
the effect of electronic strong correlation on phonon dispersion can
be observed by comparing results from LDA and LDA+\emph{U}
approaches. Compared with pure LDA calculation, the optical branches
calculated within LDA+\emph{U} shift down by around 2.5 THz and 2.3
THz for $\alpha$-UH$_{3}$ and $\beta$-UH$_{3}$, respectively.
However, the acoustic branches have no evident changes for both
phases. Due to the fact that uranium atom is much heavier than
hydrogen atom, then the optical branches denote the hydrogen
vibration while the acoustic branches come from the uranium
vibration. As a result, a large gap between optical and acoustic
modes can be observed. The key point revealed in Fig. 3 is that the
lattice dynamics behaviors are prominently influenced by electronic
strong on-site correlation, which has changed the U-H bonding
nature. This influence on lattice dynamics turns out, as analyzed
below, to be fundamental to reproduce the experimentally observed
relative thermodynamic stability sequence of $\alpha$ and $\beta$
phases of UH$_{3}$.

For the sake of determining phase transition temperature of $\alpha$-UH$_{3}$
and $\beta$-UH$_{3}$ at ambient condition, we have calculated the Helmholtz
free energy \emph{F} in the LDA and LDA+\emph{U} formalisms. This quantity at
volume \emph{V} and temperature \emph{T} can be expressed as
\begin{equation}
F(V,T)=E(V)+F_{\mathrm{{vib}}}(V,T)+F_{\mathrm{{ele}}}(V,T),\label{helmhltz}%
\end{equation}
where \emph{E}(\emph{V}) is the 0 K band energy, \emph{F$_{\mathrm{{vib}}}$%
}(\emph{V,T}) is the phonon vibrational free energy, and
\emph{F$_{\mathrm{{ele}}}$} is the thermal electronic contribution to the free
energy. The phonon vibrational free energy in the quasiharmonic approximation
can be calculated from phonon DOS $g(\omega,V)$ as $F_{\mathrm{vib}}%
(V,T)$=$k_{\mathrm{B}}T\int_{0}^{\infty}g(\omega,V)\ln[2$sinh$(\frac
{\hbar\omega}{2k_{B}T})]d\omega$. The thermal electronic contribution can be
described as $F_{\mathrm{{ele}}}(V,T)$=$E_{\mathrm{{ele}}}$$-TS_{ele}$ with
electronic entropy $S_{ele}(V,T)$=$-k_{B}\int n(\varepsilon,V)[f\ln
f$+$(1\mathtt{-}f)\ln(1\mathtt{-}f)]d\varepsilon$, where \emph{f} is
Fermi-Dirac distribution and \emph{n} is the electronic DOS. The chemical
potential at temperature \emph{T} is fixed by conservation of total valence
electron number during thermal excitation. It turns out that the value of the
thermal electronic contribution is very small, only one-tenth of the phonon
vibrational free energy. \begin{figure}[ptb]
\begin{center}
\includegraphics[width=0.8\linewidth]{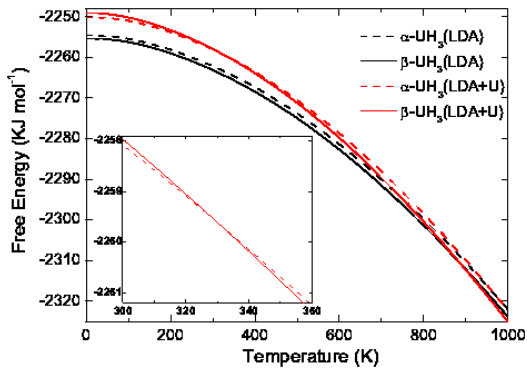}
\end{center}
\caption{(Color online) Temperature dependences of the Helmholtz free energies
within LDA (the black curves) and LDA+\emph{U} (the red/grey curves)
formalisms. The inset gives closer view of the crossing point within the
LDA+\emph{U}.}%
\label{helmholtz}%
\end{figure}

We have calculated and plotted the temperature dependences of the
Helmhotz free energy for the two UH$_{3}$ phases within LDA and
LDA+\emph{U} formalisms (Fig. \ref{helmholtz}). A significant
difference can be clearly observed between the two formalisms.
Without including the on-site Coulomb interaction, the Helmhotz free
energies of two UH$_{3}$ phases have no crossing point, and the
$\alpha$ phase is always more thermodynamically stable than the
$\beta$ phase in a wide temperature range of 0 to 1000 K, as shown
in Fig. \ref{helmholtz}. This is contrary to the experimental
observations. Whereas, after including the electronic strong
correlation, the calculated Helmhotz free energy curves of the two
allotropes cross at the temperature of 332 K, which is well
consistent with the experimental value for $\alpha
\mathtt{\rightarrow}\beta$ phase transition. Therefore, we arrive at
that while the pure LDA calculation totally fails to reproduce the
experimentally determined thermodynamic phase transition of
UH$_{3}$, the LDA+\emph{U} calculation can well depict it. This is a
good news for the science of actinide hydrides. Needless to say,
more advanced many-body techniques, such as DMFT, which can account
for some spin fluctuations at finite temperatures, will further
improve the calculations and we would like to leave it for future
consideration.

In conclusion, the ground-state properties of two allotropes of UH$_{3}$ have
been comparatively studied within the LDA and LDA+\emph{U} formalisms. After
switching on the on-site Coulomb interaction, the uranium 5$f$ states split
into the multiple well-resolved peaks that shift upward or downward with
respect to the Fermi energy, which reduces the metallicity of the 5$f$ states
and enhances the ionicity of the hydrogen bond in the system. As a result, the
calculated optical branches in the phonon dispersion of two allotropes have
been found to shift down by about 2.5 and 2.3 THz for $\alpha$-UH$_{3}$ and
$\beta$-UH$_{3}$, respectively, by turning on the Hubbard parameter. Our
theoretical Helmhotz free energy curves have shown that while the pure LDA
fails to describe the $\alpha\mathtt{\rightarrow}\beta$ phase transition, the
LDA+\emph{U} calculation gives the transition temperature of 332 K, which well
lies within the experimentally observed range. These results clearly indicate
that the electronic strong on-site correlation plays an important role in the
uranium hydride systems.

This work was supported by NSFC under Grant No. 51071032, by the Foundation
for Development of Science and Technology of China Academy of Engineering
Physics under Grant No. 2009A0102005, and by the National Basic Security
Research Program of China.

\end{document}